\documentclass[10pt,prl,twocolumn,showpacs,aps,floats,floatfix,superscriptaddress]{revtex4}

\usepackage{latexsym}
\usepackage{amsfonts}
\usepackage{amsmath}
\usepackage{amssymb}
\usepackage{amstext}

\usepackage{graphicx}
\usepackage{epsfig}
\usepackage{subfigure}
\usepackage{upgreek}
\usepackage{xcolor}

\begin{document}
\title{Disentangling collective trends from local dynamics}

\author{Marc Barth{\'e}lemy$^{\star,\#}\footnote{Corresponding author}$, Jean-Pierre Nadal$^{\#,\dagger}$, Henri Berestycki$^{\#}$}
\affiliation{$^{\star}$ Institut de Physique Th\'eorique,
CEA, IPhT CNRS, URA 2306 F-91191 Gif-sur-Yvette
France}
\affiliation{$^{\#}$ Centre d'Analyse et de Math\'ematique Sociales
  (CAMS, UMR 8557 CNRS-EHESS), Ecole des Hautes Etudes en Sciences
  Sociales, 54 bd. Raspail, F-75270 Paris Cedex 06, France}
\affiliation{$^{\dagger}$ Laboratoire de Physique Statistique (LPS,
  UMR 8550 CNRS ENS Paris 6 \& Paris 7), Ecole Normale Sup\'erieure
  (ENS), Paris, France}

\date{\today}

\begin{abstract}

  A single social phenomenon (such as crime, unemployment or birth
  rate) can be observed through temporal series corresponding to units
  at different levels (cities, regions, countries...). Units at a
  given local level may follow a collective trend imposed by external
  conditions, but also may display fluctuations of purely local
  origin. The local behavior is usually computed as the difference
  between the local data and a global average (e.g. a national
  average), a view point which can be very misleading. We propose here
  a method for separating the local dynamics from the global trend in
  a collection of correlated time series. We take an independent
  component analysis approach in which we do not assume a small
  unbiased local contribution in contrast with previously proposed
  methods. We first test our method on synthetic series generated by
  correlated random walkers. We then consider crime rate series (in
  the US and France) and the evolution of obesity rate in the US,
  which are two important examples of societal measures. For crime
  rates, the separation between global and local policies is a major
  subject of debate. For the US, we observe large fluctuations in the
  transition period of mid-$70$'s during which crime rates increased
  significantly, whereas since the $80$'s, the state crime rates are
  governed by external factors and the importance of local
  specificities being decreasing. In the case of obesity, our method
  shows that external factors dominate the evolution of obesity since
  2000, and that different states can have different dynamical
  behavior even if their obesity prevalence is similar.

\end{abstract}

\maketitle

\smallskip
{\bf Classification}: Physical sciences (Applied mathematics, Physics), Social sciences.

\smallskip 
{\bf Keywords}: Time series analysis, independent component analysis,
financial time series, crime rates, obesity.

\section{Introduction} 

Large complex systems are composed of various interconnected
components. The measure of the behavior of a single component thus
results from the superimposition of different factors acting at
different levels. Common factors such as global trends or external
socio-economical conditions obviously play a role but usually
different sub-units (such as users in the Internet, states or regions
in a country) will react in different ways and add their local
dynamics to the collective pattern. For example, the number of
downloads on a website depends on factors such as the time of the day
but one can also observe fluctuations from a user to another
one~\cite{Huberman}. In the case of criminality, favorable
socio-economical conditions will impose a global decreasing trend
while local policies will affect the regional time series. In the case
of financial series, the market imposes its own trend and some stocks
respond to it more or less dramatically. In all these cases it is
important to be able to distinguish if the stocks or regions are at
the source of their fluctuations or if on the opposite, they just
follow the collective trend.

Extracting local effects in a collection of time series is thus a
crucial problem in assessing the efficiency of local policies and more
generally, for the understanding of the causes of fluctuations. This
problem is very general and as the availability of data is always
increasing particularly in social sciences, it becomes always more
important for the modeling \cite{Castellano:2008} and the
understanding of these systems. There is obviously a huge literature
on studying stochastic signals~\cite{Kautz} ranging from standard
methods to more recents ones such as the detrended fluctuation
analysis \cite{Peng}, independent component analysis~\cite{ica}, and
separation of external and internal
variables~\cite{Menezes:2004a,Menezes:2004b}. Most of these methods
treat the internal dynamics as a small local perturbation with zero
mean which is in contrast with the method proposed here.

In a first part we present the method. In a second part, we test it on
synthetic series generated by correlated random walkers. We then apply
the method to empirical data of crime rates in the US and France, and
obesity rates in the US, for which, to our knowledge, no general
quantitative method is known to provide such separation between global
and local trends.

\section{Model and Method} 

In general, one has a set of time series $\{f_i\}_{i=1,\dots,N}(t)$
where $t=1,\dots,T$ and we will assume that the number $N$ of units
is large. The index $i$ refers to a particular unit on a specific
scale such as a region, city, a country. The problem we address
consists in extracting the collective trend and the effect of local
contributions. One way to do so is to assume the signal $f_i(t)$ to be
of the form
\begin{equation}
f_i(t)=f_i^{ext}(t)+f_i^{int}(t)
\end{equation}
where the `external' part, $f_i^{ext}(t)$, represents the impact on
the region $i$ of a global trend, while the `internal' part,
$f_i^{int}$, represents the contribution due to purely local factors.
Usually, in order to discuss the impact of local policies, one
compares a regional (local) curve $f_i$ to the average (the national
average in case of regions of a country) computed as
\begin{equation}
f^{av}(t)= (1/N) \sum_i f_i 
\label{eq:fnat}
\end{equation}
(or $f^{av}= \sum_i n_i f_i/\sum_i n_i$
if one has intensive variables and populations $n_i$). Although
reasonable at first sight, this assumes that the local component is
purely additive: $f_i(t)=f^{av}(t) +$ {\it local term}.  In this
article, following \cite{Menezes:2004a,Menezes:2004b}, we will rather
consider the possibility of having both multiplicative and additive
contributions. More specifically, we assume
\begin{equation}
f_i^{ext}(t)=a_i \; w(t)
\label{eq:fext}
\end{equation}
where $w(t)$ is a collective trend common to all series, and which
affects each region $i$ with a corresponding prefactor $a_i$. These
coefficients are assumed to depend weakly on the period considered,
ie. to vary slowly with time. We thus write
\begin{equation}
f_i(t)=a_i \; w(t) \; + \; f_i^{int}(t)
 \label{eq:hyp}
\end{equation}
We first note that the global trend $w$ is known up
to a multiplicative factor only (one cannot distinguish $a_i w$ from
$(a_i z) (w/z)$ whatever $z \neq 0$) and we will come back to this issue
of scale later. Also, the purely additive case is recovered
if the $a_i$'s are independent of $i$. If on the contrary the $a_i$'s
are different from one region to the other, the national
average (\ref{eq:fnat}), $f^{av}=\overline{f}=(1/N)\sum_if_i$,
is then given by
\begin{equation}
\overline{f}(t)=\overline{a} w(t)+\overline{f^{int}}
 \label{eq:national}
\end{equation}
Here and in the following we denote the sample average, that is the
average over all units $i$, by a bar, $\overline\cdot$, and the
temporal average by brackets $\langle\cdot\rangle$. The `naive' local
contribution is then estimated by the difference with the national
average
\begin{eqnarray}
\nonumber
f_i^{int,n}(t)&=&f_i(t)-\overline{f}(t)\\
   &=&(a_i-\overline{a}) w(t)+f^{int}_i(t)-\overline{f^{int}}(t)
\label{eq:naive}
\end{eqnarray}
The estimated local contribution $f_i^{int,n}(t)$ can thus be very
different from the original one, $f^{int}_i(t)$, and the difference
$|f_i^{int,n}(t)-f^{int}_i(t)|$ will be very large at all times $t$ where
$w(t)$ is large (note that the conclusion would be the same by taking
the national average as $f^{av}(t)= \sum_i n_i f_i/\sum_i n_i$). This
demonstrates that comparing local time series with the naive average
could in general be very misleading. Beside the correct computation of
the external and internal contributions, the existence of both
multiplicative and additive local contributions implies that the
effect of local policies must be analyzed by considering both how the
local unit $i$ follows the global trend ($a_i$) and how evolves the
purely internal contribution ($f^{int}_i$).

In a previous study~\cite{Menezes:2004a}, Menezes and Barabasi
proposed a simple method to separate the two contributions, internal
($f_i^{int}$) and external ($f_i^{ext}$ written as $a_i w(t)$).  They
assume that the temporal average $\langle f_i^{int}\rangle$ is zero,
and compute the external and internal parts by writing 
\begin{equation}
a_i=\frac{\sum_tf_i(t)}{\frac{1}{N}\sum_t\sum_j f_j(t)} = <f_i>/<\overline{f}>
\end{equation}
and $f_i^{ext}(t)=a_i \overline{f}(t)$.  This method can be shown to be correct in
very specific situations, such as the case where $f_i$ is the
fluctuating number of random walkers at node $i$ in a network, but in
many cases however, one can expect that the local contributions have a
non zero sample average 
and the method of \cite{Menezes:2004a,Menezes:2004b} will yield
incorrect results.  Indeed, if the hypothesis Eq.~(\ref{eq:hyp}) is exact,
this method would give for $w$ the estimate
$\widehat{w}(t)=\overline{a} w(t)+\overline{f^{int}}(t)$, and in the
limit $|w(t)|\to\infty$ for $t\to\infty$ would lead to the estimates
$\widehat{a}_i\approx a_i/\overline{a}$ and
$\widehat{f_i^{int}}\approx
f^{int}_i-a_i\overline{f^{int}}/\overline{a}$, which are different
from the exact results, except if $\overline{f^{int}}=0$.

In order to separate the two contributions we propose in this article
a totally different approach, by taking an independent component
analysis point of view in which we do not assume that the local
contribution has a zero average (over time and/or over the regions).
To express the idea that the `internal' contribution is by definition
what is specifically independent of the global trend, and that the
correlations between regions exist essentially only through their
dependence in the global trend, we impose that the global trend is
statistically independent from local fluctuations
\begin{equation}
\langle wf_i^{int}\rangle_c=0
\label{eq:<wfint>}
\end{equation}
(we denote by $<.>_c$ the connected correlation $\langle
AB\rangle_c=\langle AB\rangle-\langle A\rangle\langle B\rangle$), and
that these local fluctuations are essentially independent from region
to region, that is for $i \neq j$
\begin{equation}
\langle f_i^{int}f_j^{int}\rangle_c \approx 0
\label{eq:<gigj>}
\end{equation}
where this statement will be made more precise below.
We show that, for large $N$, these constraints (\ref{eq:<wfint>}),
(\ref{eq:<gigj>}) are sufficient to extract estimates of the global
trend $w$ and of the $a_i$'s.

We denote by $\mu_{w}$ the average of $w$ and by $\sigma_w$ its
dispersion, so that we write
\begin{equation}
w(t) = \mu_{w} + \sigma_w W(t)
\label{eq:wmusig}
\end{equation} 
with $\langle W\rangle=0$ and $\langle W^2\rangle=1$.
If we denote by
$F_i(t)=f_i(t)-\langle f_i\rangle$ and $G_i=f_i^{int}-\langle f_i^{int}\rangle$, 
we have
\begin{equation}
F_i(t)=A_iW(t)+G_i(t)
\end{equation}
with 
\begin{equation}
A_i=a_i\sigma_w.
\label{eq:Aidef}
\end{equation}
Note that $(\sigma_i^{ext})^2 \equiv \langle (f_i^{ext})^2\rangle_c =A_i^2$.
If we now consider the
correlations between these centered quantities, $C_{ij}=\langle F_iF_j\rangle$, we find
\begin{equation}
C_{ij}=A_iA_j+\langle G_iG_j\rangle
\end{equation}
If we assume that for $i \neq j$ $<G_iG_j>$ is negligible (of order $1/N$) compared to
$A_iA_j$ (which is what we mean by having small correlations between
internal components, Eq.~(\ref{eq:<gigj>})), from this last expression
we can show that at the dominant order in $N$, we have
\begin{eqnarray}
\sum_{j/j\neq i}C_{ij} &\simeq & A_iN\overline{A}\\
\sum_{i,j/i\neq j}C_{ij} &\simeq & N^2\overline{A}^2
\end{eqnarray}
These equations lead to
\begin{equation}
A_i=
\frac
{\sum_{j/j\neq i}C_{ij}}
{\left(\sum_{j,j'/j \neq j'}C_{jj'}\right)^{1/2}}
\label{eq:Ai}
\end{equation}
which is valid when $\langle \overline{G}^2\rangle\ll
\overline{A}^2$. We note that our method has a meaning only if strong
correlations exist between the different $f_i$'s and if it is not the
case, the definition of a global trend makes no sense and the
approximation used in our calculations are not valid.

In the Supporting Information (section SI1) we show that the factors
$A_i$'s can also be computed as the components of the eigenvector
corresponding to the largest eigenvalue of $C_{ij}$ - a method which
is valid under the weaker assumption of having a small number
(compared to $N$) of non diagonal terms of the matrix $D_{ij}=\langle
G_iG_j\rangle$ which are not negligible.

Once the quantities $A_i$ are known, we can compute the global
normalized pattern $W(t)$ with the reasonable estimator given by
$\overline{F/A}$,
\begin{equation}
W(t) \simeq \overline{\frac{F}{A}}
\label{eq:W}
\end{equation}
Indeed,
\begin{equation}
\overline{\frac{F}{A}}(t) = \frac{1}{N}\sum_i\frac{F_i}{A_i}=W(t)+\overline{\frac{G}{A}}
\label{eq:FA}
\end{equation}
and since 
the quantity $\overline{G/A}$ is a sum of independent variables
with zero mean, we can expect it to behave as $1/\sqrt{N}$. 
We can
show that this actually results from the initial assumptions. Indeed,
by construction $\langle \overline{G/A}\rangle =0$ and the second moment is
\begin{equation}
\langle\left(\overline{\frac{G}{A}}\right)^2\rangle=\frac{1}{N^2}
\sum_{ij}\frac{\langle G_iG_j\rangle}{A_iA_j}
\end{equation}
By assumption we have $\langle G_iG_j\rangle\approx 0$ if $i\neq j$ and we thus 
obtain $\overline{G/A}\sim 1/\sqrt{N}$.

The computation of the $A_i$'s and of $W$ is equivalent to an
independent component analysis (ICA) \cite{ica} with a {\it single}
source (the global trend) and a large number $N$ of sensors. However,
in contrast with the standard ICA, we are not interested in getting
only the sources (here the trend $W$), but also the internal
contributions (which, in a standard ICA framework, would be considered
as noise terms, typically assumed to be small). We have already the
$A_i$'s, and since $W(t)$ has been calculated we can compute
$G_i=F_i(t)-A_i W(t)$. We thus obtain at this stage
\begin{equation}
\langle f_i \rangle = A_i \frac{\mu_{w}}{\sigma_w} + \langle f_i^{int}\rangle.
\label{eq:mui}
\end{equation}
This is a set of $N$ equations 
for $N+1$ unknown ($\mu_{w}/\sigma_w$ and the $\langle
f_i^{int}\rangle$'s) and we are thus left with one free parameter, the
ratio $\mu_{w}/\sigma_w$. Knowing its value would give the $N$ local
averages, the $\langle f_i^{int}\rangle$'s.  Less importantly one may
want also to fix the average $\mu_{w}$ (hence both $\mu_{w}$ and
$\sigma_w$) in order to fully determine the pattern $w(t)$: this will
be of interest only for making a direct comparison between this
pattern and the national average (\ref{eq:fnat}).  This
equation~(\ref{eq:mui}) suggests a statistical linear correlation
between $\langle f_i \rangle$ and $A_i $, with a slope given by
$\mu_{w}/\sigma_w$. We will indeed observe a linear correlation in the
data sets (next section, Figure~$2$). However, it could
be that the $\langle f_i^{int}\rangle$'s themselves are correlated
with the $A_i$s. Hence, and unfortunately, a linear regression cannot
be used to get an unbiased estimate of the parameter
$\mu_{w}/\sigma_w$. In the absence of additional information or
hypothesis this parameter remains arbitrary. However one may compare
the qualitative results obtained for different choices of
$\mu_{w}/\sigma_w$: which properties are robust, and which ones are
fragile. In particular one would like to be able to access how a given
region is behaving, compared to another given region, and/or to the
global trend. To do so, in the applications below we will in
particular analyze: (i) the correlations between the two local terms,
$A_i$ and $\langle f_i^{int}\rangle$; (ii) the robustness of the rank
given by the $\langle f_i^{int}\rangle$'s; (iii) the sign of $\langle
f_i^{int}\rangle$; (iv) the quantitative and qualitative similarities
between $f_i^{int}(t)$ and the naive estimate $f_i^{int,n}(t)$.

We will focus on two particular scenarios. First, one may ask the
global trend to fall `right in the middle' of the $N$ series.  There
are different ways to quantify this.  One way to do so is to note
that, in the absence of internal contribution, $f_i/a_i$ would be
equal to $w$, hence $\langle f_i \rangle/A_i$ would be equal to
$\mu_{w}/\sigma_w$. Therefore we may compute $\mu_{w}/\sigma_w$ by
imposing
\begin{equation}
\frac{\mu_{w}}{\sigma_w} = \frac{1}{N}\sum_i\frac{\langle f_i\rangle}{A_i},
\label{eq:gamma}
\end{equation}
which is thus equivalent to impose $ \frac{1}{N}\sum_i\frac{\langle
  f_i^{int}\rangle}{A_i} = 0.$ An alternative is to ask the resulting
$f_i^{int}$ to be as close as possible to the naive ones
(Eq.~(\ref{eq:naive})), by minimizing $(1/N)\sum_i
<(f_i^{int}-f_i^{int,n})^2>$ which gives
\begin{equation}
\frac{\mu_{w}}{\sigma_w} = \frac{ \langle f^{av} \rangle \overline{A} }{\overline{A^2}}
\label{eq:alt2}
\end{equation}
In both cases one may then fix $\mu_{w}$ from $\mu_{w}=\langle
f^{av}\rangle$ or by imposing $w(t_0)=f^{av}(t_0)$ for some arbitrary
chosen $t_0$. Finally, one may rather ask for a conservative
comparison with the naive approach by minimizing the difference
between $w$ and $f^{av}$: either by writing $\mu_{w}=\langle
f^{av}\rangle$ (or $w(t_0)=f^{av}(t_0)$) and $\sigma_w=\langle
(f^{av})^2\rangle_c$, or by minimizing $\langle(w-f^{av})^2\rangle$,
which gives
\begin{equation}
\mu_{w} = \langle f^{av}\rangle \;\;\;\mbox{and}\;\;\;
\sigma_w  =  \langle W f^{av} \rangle
\label{eq:alt4}
\end{equation}
For $N$ is large, one can check that the results depend weakly on any
one of these reasonable choices.

The second scenario considers the correlations between the $\langle
f_i^{int}\rangle$'s and the $A_i$'s.  As we will see, the first
hypothesis leads to a strictly negative correlation.  An alternative
is thus to explore the consequences of assuming no correlations, hence
asking for
\begin{equation}
\overline{ A\; \langle f^{int}\rangle} \;-\;  \overline{A} \; \overline{\langle f^{int}\rangle} \;=\; 0 
\label{eq:corga}
\end{equation}
which implies that the slope of the observed linear correlation $\langle f_i\rangle$ with $A_i$ gives the value of $\mu_{w}/\sigma_w$.
As explained above, for each application below we will discuss the robustness of the results with respect to these choices of the parameter $\mu_{w}/\sigma_w$.

\medskip 

We can now summarize our method. It consists in (i) estimating the
$A_i$'s using Eq. (\ref{eq:Ai}) (or using the eigenvector
corresponding to the largest eigenvalue of the correlation matrix,
section SI1), (ii) computing $W$ using Eq.~(\ref{eq:W}), and finally
(iii) comparing the results for different hypothesis on
$\mu_{w}/\sigma_w$ as discussed above.  We propose to call this
method 
the {\it External Trend and Internal Component Analysis} ({\bf
  ETICA}).  We note that if the hypothesis Eq.~(\ref{eq:hyp}),
(\ref{eq:<wfint>}), (\ref{eq:<gigj>}) are correct, the method gives
estimates of $W$, the $A_i$s (hence of $f_i^{int}-\langle
f_i^{int}\rangle$) which become exact in the limit $t$ and $N$ large,
and a good estimate of the full trend $w$ (hence of the $\langle
f_i^{int}\rangle$) whenever this trend, qualitatively, does fall `in
the middle' of the time series.

Once we have extracted with this method the local contribution
$f_i^{int}$, and the collective pattern $w(t)$ together with its
redistribution factor $a_i$ for each local series, we can study
different quantities, as illustrated below on different applications
of the method.  In general, although this method gives a pattern
$w(t)$ very similar to the sample average $\overline{f}(t)$, we will see
that there is non trivial structure in the prefactors $a_i$'s leading
to non trivial local contributions $f_i^{int}(t)$.

In some cases one may expect to have, in addition to the local
contribution, a linear combination of several global trends (a small
number of 'sources'): we leave for future work the extension of our
method to several external trends.

\section{Applications: correlated random walkers, crime rates in
  the US and France, obesity in the US.}

We first test our method on synthetic series and we then illustrate
it on crime rate series (in the United States and in France) and on
US obesity rate series. For the crime rates, a plot of the time series
shows that obviously a common trend exists (Fig. 1).
\begin{figure}[!h]
\begin{tabular}{c}
\epsfig{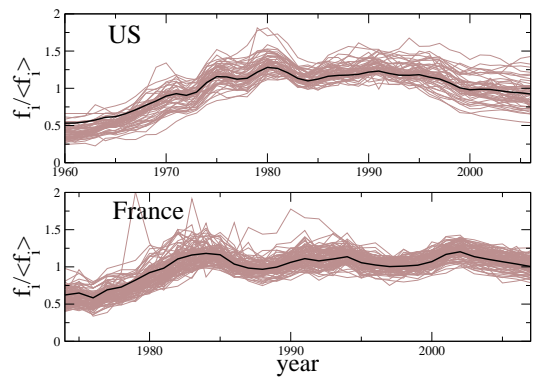}\\
\end{tabular}
\caption{ Collective pattern. Crime rates for the US (upper panel) and France (lower
panel) normalized by their time average. The black thick line
represent the collective pattern $w(t)$ computed with our method.}
\end{figure}
After computing the internal and external terms, we perform different
tests in order to assess the validity of the approach. In particular,
Figure~$2$ shows a plot of the local factors $A_i$s versus the data
time-averages, the $\langle f_i\rangle$'s. 
\begin{figure}[!h]
\begin{tabular}{c}
\epsfig{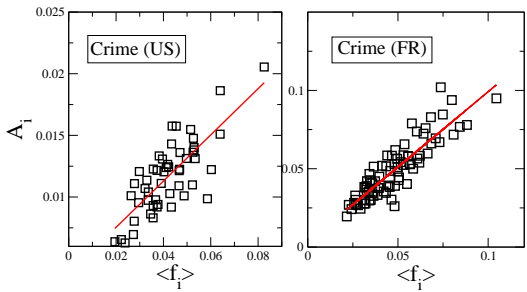}\\
\end{tabular}
\caption{ Existence of a linear correlation. We plot the
prefactors $A_i$ versus the time average $\langle f_i\rangle$ for the
three different datasets.}
\end{figure}
One observes a statistical linear correlation in the four set of time
series. We stress that the $A_i$'s are computed from the covariance
matrix of the data, hence after removing the means from the time
series.  The fact that we do observe a linear correlation is thus a
hint that our hypothesis on the data structure is reasonable (in
contrast the very good linear correlation observed in
\cite{Menezes:2004a,Menezes:2004b} can be shown to be an artefact of
the method used in these works, leading to an exact proportionality
independently of the data structure, (see the section SI2).  We now
discuss in more detail the synthetic series, each one of the crime
rate data sets, and the obesity rate.

\subsection{Synthetic series: correlated random walkers.}

We can illustrate our method on the case of correlated random walkers
described by the equation
\begin{equation}
f_i(t)=F(t)+\sum_{\tau=0}^t\xi_i(\tau)
\end{equation}
where $F(t)$ is the global trend imposed to all walkers and the
$\xi_i(t)$ are gaussian noises but with possible correlations between
different walkers
$\overline{\xi_i(t)\xi_j(t)}=[(N-M)\delta_{ij}+\alpha^2M]/12$ where
$\alpha$ and $M$ are tunable parameters (see the supporting section
SI3).  For $M=0$, the random noises $\xi_i(t)$ are independent and our
method is very accurate: we choose for example a sinusoidal trend
$F(t)=\sin(\omega t)$ and we plot in the figure ~$3$ the original
signal, the exact local contribution and the local contribution
computed with our method. 
\begin{figure}[!h]
\begin{tabular}{c}
\epsfig{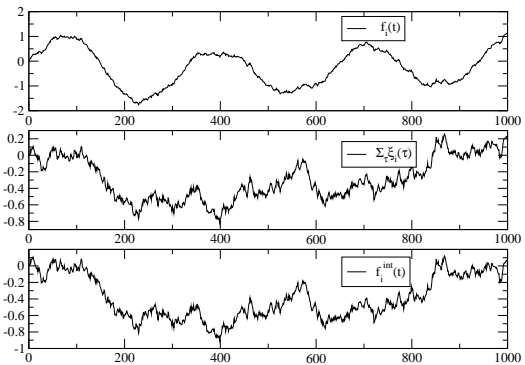}\\
\end{tabular}
\caption{ (A) Original signal composed of the superimposition of 
a sinusoidal trend and gaussian noises (for $N=100$ walkers). (B) Exact local
contribution. (C) Local contribution extracted with our method.}
\end{figure}
When the correlation between walkers is increasing we study the
Pearson correlation coefficient between the original local
contribution and the estimate provided by our method, and we observe
that our method is indeed accurate as long as the correlations between
the $G_i$'s are not too large, which corresponds here to the condition
$\alpha^2 M\ll 1$.

\subsection{Crime rates in the US and France.}  

In criminology an essential question concerns the impact of local
policies, a subject of much debate \cite{dMR04,Zimring:2007}. In order
to assess these local effects (at the level of a state or a region),
most authors consider the difference of a state evolution with the
national average. As we noticed above this may lead to incorrect
predictions.  In this second part of applications, we thus illustrate
our method on the analysis of the series of crime rates in $50$ states
in the US \cite{refcrimeUS} for the period $1965-2005$, and about
$100$ {\it d\'epartements} of France~\cite{refcrimeFR} for the period
$1974-2007$.  On Fig.~$1$ we represent these time series normalized by
their time average. The observed data collapse confirms the existence
of a collective pattern (we also show on this plot the collective
pattern $w(t)$ obtained with our method).  For the French case, we
have withdrawn outliers which do not satisfy our initial
assumptions. The series of these {\it d\'epartements} are indeed
uncorrelated with the rest of crime rates and cannot be incorporated
in the calculation of the collective pattern.  We apply our method to
these data and extract $w(t)$, the $A_i$'s and $f_i^{int}(t)$.  As
already mentioned, we plot on Fig.~$2$ the $A_i$'s vs. the averages
$\langle f_i\rangle$, exhibiting a statistical linear correlation. We
can check a posteriori that all conditions assumed in our calculation
are fulfilled (zero $\langle w f_i^{int}\rangle$ and small $\langle
G_iG_j\rangle$, see SI1). Also, we checked that the coefficients $a_i$
do not vary too much the period considered, which is an important
condition for our method (see the discussion on different datasets in
the SI4).

In order to assess quantitatively the importance of local versus external
fluctuations, we study in particular the ratio of dispersions defined
by
\begin{equation}
\eta_i=\frac{\sigma_i^{ext}}{\sigma_i^{int}}
\label{eq:eta}
\end{equation}
where the external contribution is the standard deviation of
$f_i^{ext}(t)=a_i w(t)$, that is $\sigma_i^{ext}= A_i$, and the
internal one is given by $(\sigma_i^{int})^2=\langle
(f_i^{int})^2\rangle_c = \langle G_i^2\rangle$. Note that these
quantities $\eta_i$, being based on fluctuations, does not depend on
$\mu_w/\sigma_w$.  This quantity is found in both cases in France and
in the US larger than one. This indicates that external factors always
dominate over local fluctuations, while local policies seem to play a
minor role. In the case of crime, these external effects might be
socio-economical factors such as unemployment, density, etc.

In addition to compute the average of the $\eta_i$'s, we can also observe the
time evolution of the heterogeneity defined by the sample variances of the
different components. We first observe on Fig.~$4$ that large
fluctuations are observed in the transition period of mid-$70$'s
during which crime rates increased significantly. 
\begin{figure}[!h]
\begin{tabular}{c}
\epsfig{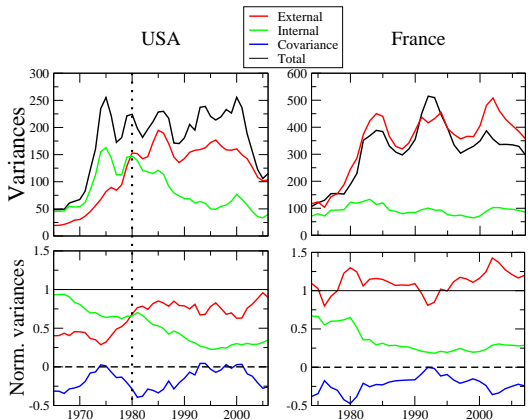}\\
\end{tabular}
\caption{ Comparison of internal and external fluctuations. On
the left (right) column we present the results for the US (France). On
the upper panels, we represent the total variance of the signal, the
external and the internal contribution. On the lower panels, we
represent the external, internal, and the covariances normalized by
the variance of the signal. We can observe that for the US, the
external contribution is dominating since the $80$'s.}
\end{figure}
We also observe for the USA that until $1980$, fluctuations were
essentially governed by local effects but that this trend is inverted
and increases in the period post-$80$'s. In particular during the
period $1980-2000$ during which one observes a decline of crime
rates~\cite{Zimring:2007}, it is the collective trend which determines
the fluctuations.

Even we have presented results for reasonable choices of the parameter
$\sigma_w$ (in the following we make the harmless choice $\mu_w=1$),
one can ask the question of the robustness of different observed
properties. First, we can compare the predictions for $\sigma_w$
obtained for the different assumptions used in this paper. In the
upper panels for Figs.~$5$ and $6$ we show for the US (France), the
quantities $\overline{\langle f_i^{int}\rangle/a_i}$,
$\overline{\langle f_i^{int}\rangle}/\overline{a}$ and
$r=(\overline{\langle f_i^{int}\rangle a}-\overline{\langle
  f_i^{int}\rangle}\overline{a})/\sigma_a^2$.
\begin{figure}[!h]
\begin{tabular}{c}
\epsfig{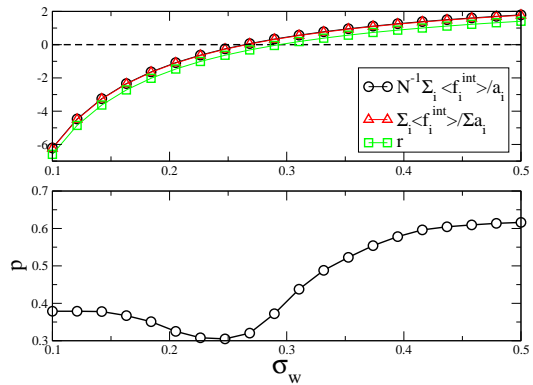}\\
\end{tabular}
\caption{ Determination of $\sigma_w$ in the US crime rate
case. We can use various conditions in order to determine $\sigma_w$:
$0=N^{-1}\sum_i<f_i^{int}>/a_i$, $0=\sum_i<f_i^{int}>/\sum_ia_i$, or
$r=0$ ($r$ is defined in the text). We see in this plot that they all
give very similar values. Lower panels: average fraction of time for
which $\langle f_i^{int}\rangle$ has the same sign as the naive
calculation $\langle f_i\rangle-\langle f_i^{av}\rangle$.}
\end{figure}
We see on these figures that these quantities are zero for values of
$\sigma_w$ which are very close. We also compute the fraction of time
$p_i$ for which $f_i^{int}(t)$ and the naive calculation $\langle
f_i\rangle-\langle f_i^{av}\rangle$ have different signs. 
\begin{figure}[!h]
\begin{tabular}{c}
\epsfig{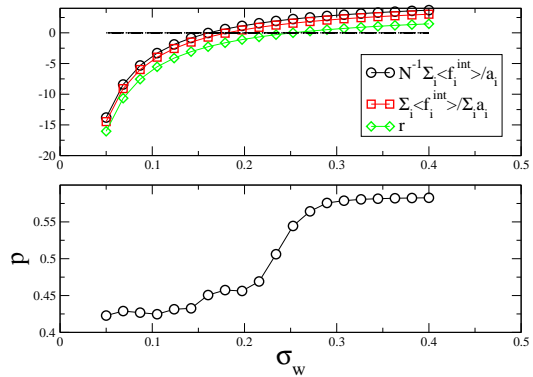}\\
\end{tabular}
\caption{ As in figure $5$, we can determine $\sigma_w$ in the
case of the crime rate in France, by using different conditions:
$0=N^{-1}\sum_i<f_i^{int}>/a_i$, or $0=\sum_i<f_i^{int}>/\sum_ia_i$,
or $r=0$. Here also, these conditions give very similar values of
$\sigma_w$. Lower panels: average fraction of time for which $\langle
f_i^{int}\rangle$ has the same sign as the naive calculation $\langle
f_i\rangle-\langle f_i^{av}\rangle$.}
\end{figure}
We plot in the lower panels of Figs.~$4$ and $5$, the quantity
$p=\frac{1}{N}\sum_ip_i$ showing a that for this range of $\sigma_w$,
the signs of $\langle f_i^{int}\rangle$ and $\langle
f_i\rangle-\langle f_i^{av}\rangle$ are the same for about $60\%$ of
the time period.  We can also study the sign $\langle
f_i^{int}\rangle$ versus $\sigma_w$ and we can observe some
robustness. In particular, in the US case, approximately $6$ states
(CA, NV, MO, MI, NY, AZ) have a positive local contribution (in the
range $\sigma_w\in [0.24,0.32]$ while $6$ states have always a
negative local contribution (VT, GA, LA, NH, CT, MS). In these cases
we can reasonably imagine that local policies have a noticeable
effect.

Finally, we can also analyze the ranking of the local contributions
$\langle f_i^{int}\rangle$ versus $\sigma_w$ by studying Kendall's
$\tau$ for the two consecutive series $\{\langle
f_i^{int}\rangle\}(\sigma_w)$ and $\{\langle
f_i^{int}\rangle\}(\sigma_w+\delta\sigma_w)$. In both cases (France
and US) we observe a $\tau$ larger than $0.9$ for the range chose
$\sigma_w\in [0,0.5]$ (the control case for a random permutation being
less than $0.1$) indicating a large robustness of the ranking. This
means that independently of the assumption used to compute $\sigma_w$
we can rank the different regions according to the importance of their
local contribution.

\subsection{Obesity in the US.}

The prevalence of obesity (defined as a body max index \-- BMI, which is 
the ratio of the body mass to the square of the height \-- larger
than $30 kg/m^2$) is rapidly increasing in the world \cite{James:2001}
and reached epidemic proportion in the US and is now a major public
health concern \cite{Mokdad:1999,Ogden:2006}. 

Disparities by sex and between ethnic groups have been observed in the
prevalence of obesity \cite{Beydoun:2007}, but few studies focus on
the effect of local factors and policies on the obesity rate. We thus
apply our method to data from the CDC \cite{nccd} which describe the
percentage of the population which is obese for each states in the US
and for the period 1995-2008. As in the crime rate case, we can
compare the variances for the internal and external contributions (see
SI5) and we observe that the external contribution is dominating since
the year $2000$. This result means that the global trend is the major
cause of the evolution of obesity in different states. We can get more
detailed information about the specific behavior of the states by
studying the ratio $\eta_i$ defined in Eq. (26) and the ratio of the
fraction of the time average local contribution to the total signal
$y_i=\langle f_i^{int}\rangle/\langle f_i\rangle$.  We represent these two
quantities in a plane (see figure $7$) and we first note that for all
states $\eta_i>1$ which means that fluctuations are mainly governed by
the global trend. 
\begin{figure}[!h]
\begin{tabular}{c}
\epsfig{file=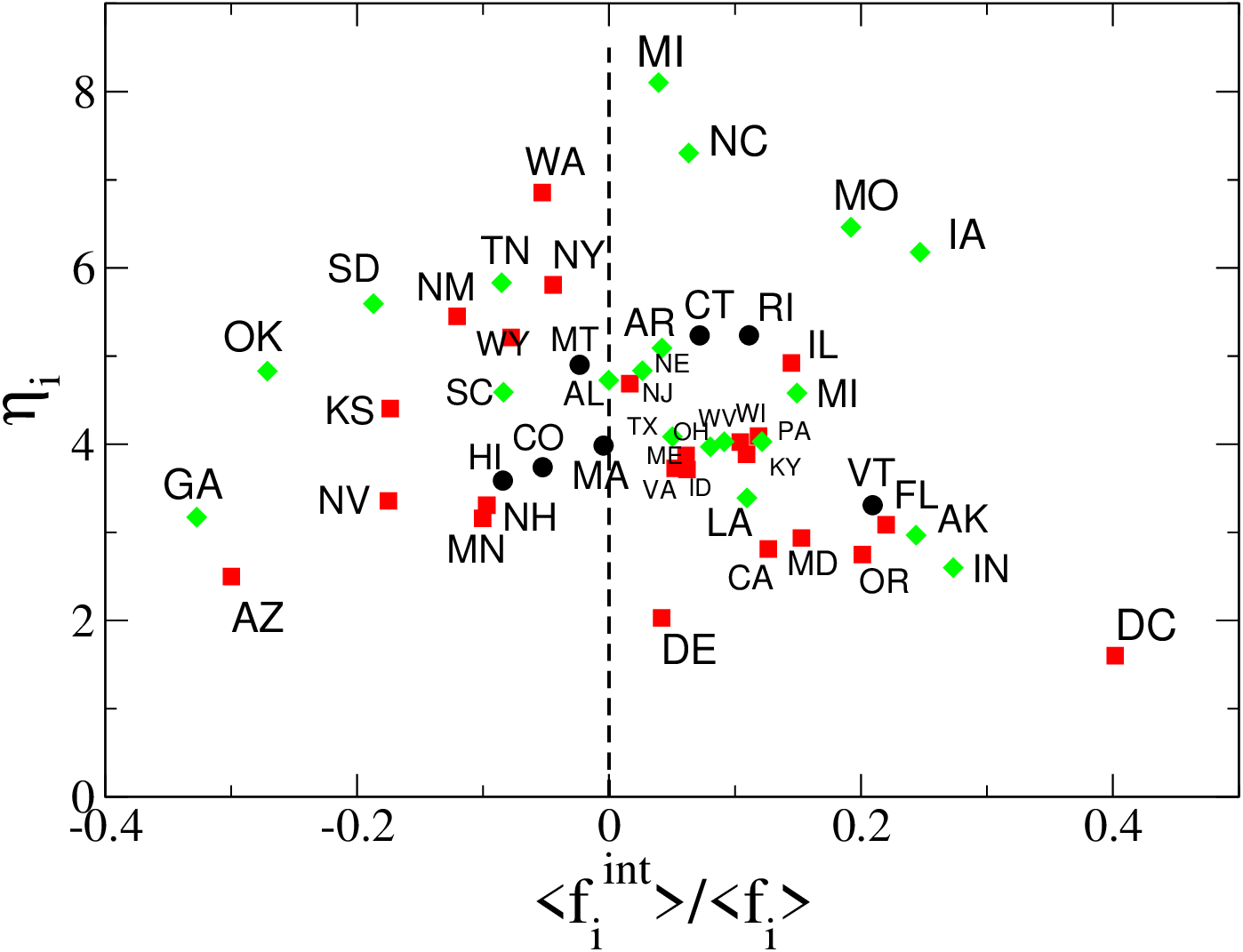,width=0.8\linewidth,clip=}\\
\end{tabular}
\caption{ Fluctuations versus importance of the local contribution. We
plot the quantity $eta_i$ versus $y_i=\langle f_i^{int}\rangle/\langle
f_i\rangle$ for the different US states. We divide the states in three
groups (circles: share less than $22\%$; squares: share in the interval
$[22\%,26\%]$; diamonds: share larger than $26\%$). Low prevalence
states seem to concentrate in the same region $y_i\approx 0$, while
medium- and large-prevalence states display very different values of
$\eta_i$ and $y_i$.}
\end{figure}
We can also divide the states into two groups (with
$y_i>0$ and $y_i<0$). For large and positive $y_i$, the states have a
small $a_i$ which means that these states are the less susceptible to
the global trend, while in the opposite case, the states are governed
by the global trend. Within each group we can then distinguish the
states according to their level of fluctuations ($\eta_i$ close to or
much larger than one).  The states Arizona, Georgia, and Oklahoma for
example have very little local contribution and their variations is
dominated by the global trend. In this respect, states such as DC,
Indiana are very different from the first group. More generally, we
can see on this figure that states with large prevalence display very
different values of $(y_i,\eta_i)$. This result points toward the fact
that describing states by their prevalence only can be very misleading
and can hide important dynamical behaviors. Finally, we also computed
the quantities $y_i$ and $\eta_i$ using the `naive' local contribution
using the national average $f^{av}(t)$ defined in Eq.~(\ref{eq:fnat})
by $f_i^{int,n}(t)=f_i(t)-f^{av}(t)$. We represent in figure $8$ the
difference as vectors of components given by $(\langle
f_i^{int}-f_i^{int,n}\rangle/\langle f_i\rangle,\eta_i-\eta_i^{naive})$ and
we can see on this figure that for roughly half of the states the
naive calculation of the local contribution can be very misleading.
\begin{figure}[!h]
\begin{tabular}{c}
\epsfig{file=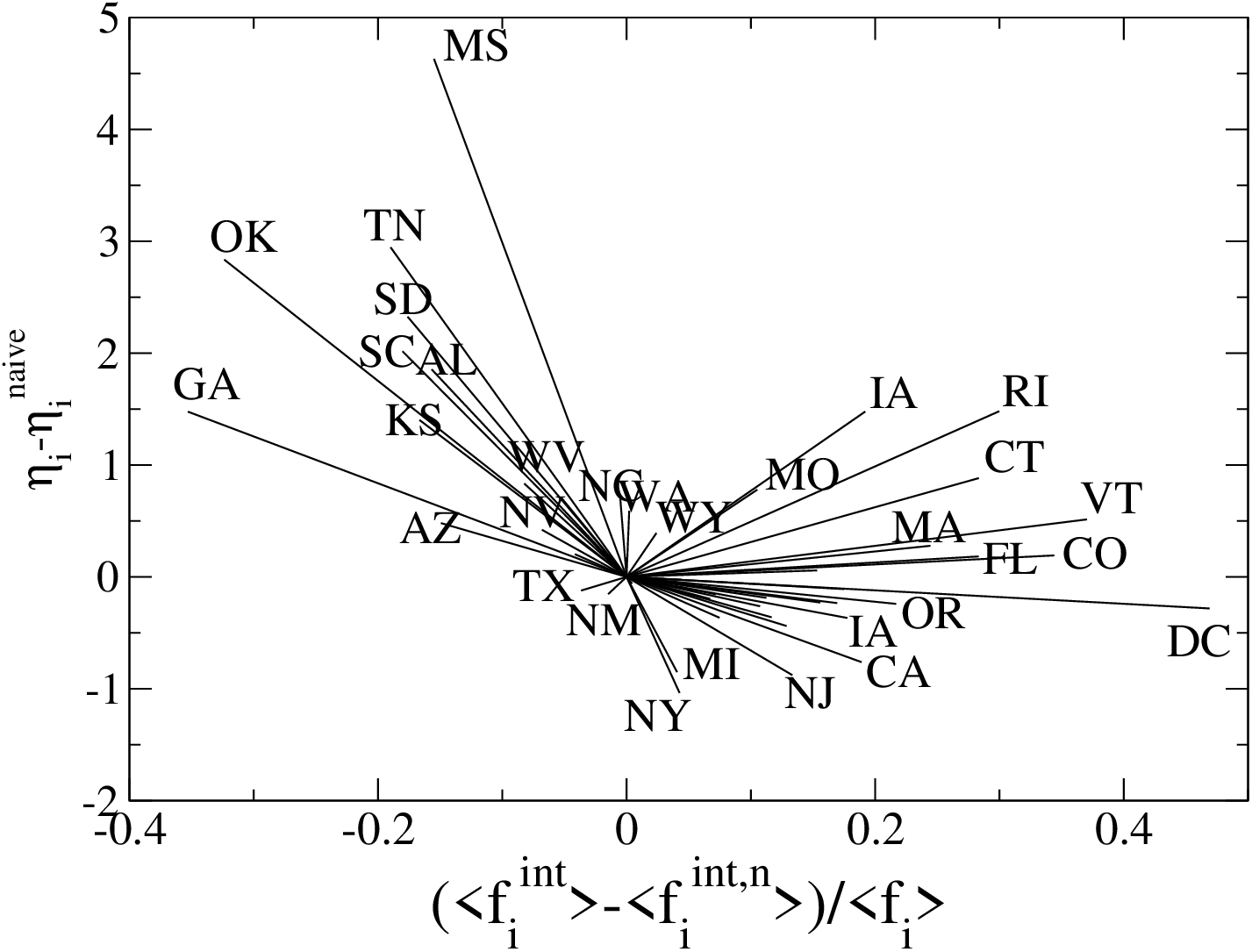,width=0.8\linewidth,clip=}\\
\end{tabular}
\caption{ Difference with the naive fluctuations and local 
contribution. We represent for the different states the difference vectors
$(\langle f_i^{int}-f_i^{int,n}\rangle/\langle
f_i\rangle,\eta_i-\eta_i^{naive})$ (for the sake of clarity, we indicated the name of the
corresponding state for most vectors except for those close to
$(0,0)$). For half of the state the difference
between the naive calculation and our method is not negligible.}
\end{figure}

\section{Discussion.}  

In this article we adressed the crucial problem of extracting the
local components of a system governed by a global trend. In this case,
comparing the local signal to the average is very misleading and can
lead to wrong conclusions. We applied this method to the example of
crime rates series in the US and France and our analysis revealed
surprising facts. The important result is about the importance of
fluctuations which after the $80$'s in the US are governed by external
factors. This result suggest that understanding the evolution of crime
rates relies mostly on the identification of global socio-economical
behavior and not on local effects such as state policies etc. In
particular, this result could also help in understanding the
decreasing trend observed in the US and which so far remains a
puzzle~\cite{Zimring:2007,Levitt}. In the case of obesity, we show
that since the year $2000$, external factors dominate, and maybe more
importantly that states with the same level of prevalence have very
different dynamical behaviors, thus calling for the need of a
detailled study state by state.

However one may expect an even better signal analysis by assuming that
there are several independent external trends: it will be interesting
to see if our approach, combined with the more standard ICA
techniques, can be generalized to the case of several global trends (a
small number of 'sources').  The recent availability of large amounts
of data in social systems call for the need of tools able to analyze
them and to extract meaningful information and we hope that our
present contribution will help in the understanding of these systems
where the local dynamics is superimposed to collective trends.

\noindent{\bf Acknowledgements}: We thank the anonymous referees for
constructive remarks, in particular about the applicability conditions
of our method. This work is part of the project
``DyXi'' supported by the French National Research Agency, the ANR
(grant ANR-08-SYSC-008).



\section{ Supporting information 1. Determining the $A_i$'s by using eigenvectors of the correlation matrix}

The data correlation matrix $C_{ij}$ is known to provide useful
information, in particular for the analysis of financial time series
$[9,10]$ or in other fields, e.g. in protein structure analysis
$[11]$. The first, largest, eigenvalue is related to a global trend,
and usually one is interested in the small number of intermediate
eigenvalues: the associated eigenvectors give the relevant
correlations in the data -- e.g. allows to extract the sectors in
financial time series. Here, making explicit use of our hypotheses, we
extract from the first eigenvector of the correlation matrix the $A_i$
factors which give how the global trend is amplified or reduced at the
local level.

We have
\begin{equation}
C_{ij}=A_iA_j+D_{ij}
\end{equation}
where $D_{ij}=\langle G_iG_j\rangle$.
If $\psi$ is a normalized eigenvector ($\psi\cdot\psi=1$) of $C$ with eigenvalue
$\lambda$: $C\cdot\psi=\lambda\psi$, we have
\begin{equation}
C\cdot\psi=(A\cdot\psi)A+D\cdot\psi
\end{equation}
We can have $A\cdot\psi$ which implies that $\psi$ is also eigenvector
for $D$ which in general is unlikely (there are no reasons that eigenvectors of $D$ are
orthogonal to $A$). If $A\cdot\psi\neq 0$ we then obtain
\begin{equation}
\lambda=A\cdot A+\frac{A\cdot D\cdot\psi}{A\cdot\psi}
\end{equation}
and
\begin{equation}
\psi=\frac{A\cdot\psi}{\lambda}A+\frac{D\cdot\psi}{\lambda}
\end{equation}
For the largest eigenvalue, we will neglect at first order the second term of the rhs of this last
equation, which leads to $\psi\propto A$. Since $\psi$ is normalized, we obtain
\begin{equation}
\psi\approx \frac{A}{\sqrt{A\cdot A}}
\end{equation}
This approximation is justified if $A\cdot D\cdot\psi$ is small compared
to $A\cdot A$ and thus
\begin{equation}
\frac{A\cdot D\cdot A}{(A^2)^2}\ll 1
\end{equation}
Since $A \cdot A={\cal O}(N)$, this approximation is justified if $A\cdot D\cdot A$ is of
order $N$ and not of order $N^2$. This is correct if $D$ is diagonal (which means 
that the external components are not correlated $\langle G_iG_j\rangle\propto\delta_{ij}$),
but also if the number of non-zero terms of $D_{ij}$ is finite compared to $N$, or in other words
if $D$ is a sparse matrix.

We compared the values of $A_i$ computed with the method exposed in the text and 
with the eigenvector method. Results are reported in the figures (9,10,11).

\begin{figure}[!h]
\begin{tabular}{c}
\epsfig{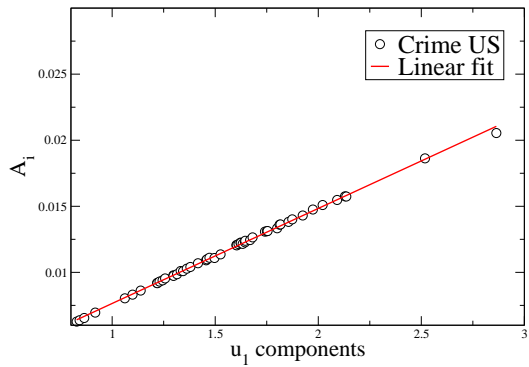}\\
\end{tabular}
\label{fig:eigen1}
\caption{Comparison of the $A_i$ computed with
  expressions in the text (Eq. $16$) and with the components of the
  eigenvector corresponding to the large eigenvalue of $C_{ij}$ in 
the case of crime rates in the US.}
\end{figure}

\begin{figure}[!h]
\begin{tabular}{c}
\epsfig{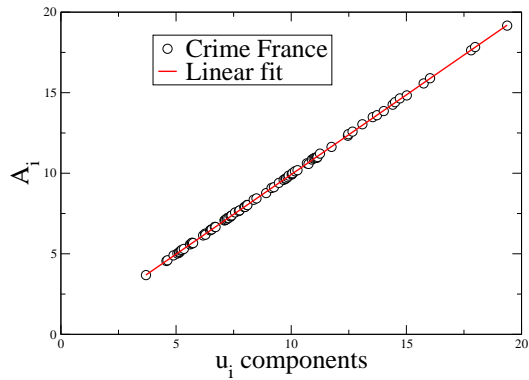}\\
\end{tabular}
\label{fig:eigen2}
\caption{Comparison of the $A_i$ computed with
  expressions in the text (Eq. $16$) and with the components of the
  eigenvector corresponding to the large eigenvalue of $C_{ij}$ in 
the case of crime rates in France.}
\end{figure}

\begin{figure}[!h]
\begin{tabular}{c}
\epsfig{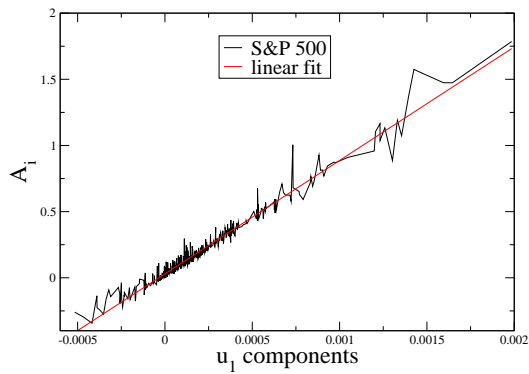}\\
\end{tabular}
\label{fig:eigen3}
\caption{Comparison of the $A_i$ computed with
  expressions in the text (Eq. $16$) and with the components of the
  eigenvector corresponding to the large eigenvalue of $C_{ij}$ in the
case of the $S\& P500$.}
\end{figure}

We see that indeed for the crime rates in the US and in France, $D_{ij}$ is indeed negligible which
demonstrate that the correlations of the internal contributions between different
states in the US are negligible. This is not the case for the stocks
in the $S\& P500$ where we can observe (small) discrepancies between the
two methods, a result which supports the idea of sectors in the $S\& P500$.

\section{ Supporting information 2: Scaling}

We show that the scaling $\sigma_i^{ext} \sim <f_i>$ observed by de
Menezes and Barabasi in $[6,7]$ is actually built in the method
proposed by these authors: it is a direct consequence of their
definitions of the internal and external parts, and it does not depend
on the data structure.

Indeed, let $f_i(t), t=1,...,T, i=1,...,N$ be an arbitrary data set
such that $<\bar f> \neq 0$. For $i=1,...,N$, following $[6]$ define $A_i^{MB}$ by
\begin{equation}
A_i^{MB}  \equiv \frac{<f_i>}{<\bar f>} 
\label{eq:AiB}
\end{equation}
and $f_i^{MB,ext}(t)$ by
\begin{equation}
f_i^{MB,ext}(t) \equiv A_i^{MB} \bar f (t)
\label{eq:fextB}
\end{equation}
Then, from these definitions and without any hypothesis or constraint on the data other than $<\bar f> \neq 0$, one has
\begin{equation}
<f_i^{MB,ext}> = A_i^{MB} <\bar f (t) > = <f_i>
\label{eq:<fiext>}
\end{equation}
and
\begin{equation}
<(f_i^{MB,ext})^2> = (A_i^{MB})^2 <\bar f (t)^2 >.
\label{eq:<fiext2>}
\end{equation}
Hence
\begin{equation}
(\sigma_i^{MB,ext})^2 = (A_i^{MB})^2 \; \sigma_f^2 = <f_i>^2\; \frac{\sigma_f^2}{<\bar f>^2} 
\label{eq:sigext}
\end{equation}
with
\begin{equation}
\sigma_f^2  \equiv <\bar f (t)^2 > - <\bar f (t) >^2
\label{eq:sigf}
\end{equation}
Hence, one has always
\begin{equation}
\sigma_i^{MB,ext} = \frac{\sigma_f}{|<\bar f>|}  |<f_i>|
\label{eq:always}
\end{equation}
The dispersion of the external component, if defined from
(\ref{eq:AiB}) and (\ref{eq:fextB}), is thus {\it exactly}
proportionnal to the mean value of the local data.

\section{ Supporting information 3. Synthetic series: correlated random walkers}

We considered the case where the external trend is 
\begin{equation}
F(t)=\sin(\omega t)
\end{equation}
The gaussian noises are given by
\begin{equation}
\xi_i(t)=\alpha\sum_{j=1}^{M}u_j^{(0)}(t)+\sum_{j=M+1}^Nu_j^{(i)}(t)
\end{equation}
where the $u_j^{(0)}(t)$ and $u_j^{(i)}(t)$ are independent, uniform
random variable of zero mean and variance equal to $1/12$. In this
case, the correlation between different noises are governed by the
parameters $\alpha$ and $M$
\begin{equation}
\overline{\xi_i\xi_j}=\frac{\alpha^2M}{12}+\frac{N-M}{12}\delta_{ij}
\end{equation}
When $M=0$, the variables $\xi_i$ and $\xi_j$ are independent (for
$i\neq j$) and we can monitor the correlations by increasing the value
of $M$. We plot in figure~$12$, $N=100$ random walkers
in the usual uncorrelated case and in presence of correlations.
\begin{figure}[!htp]
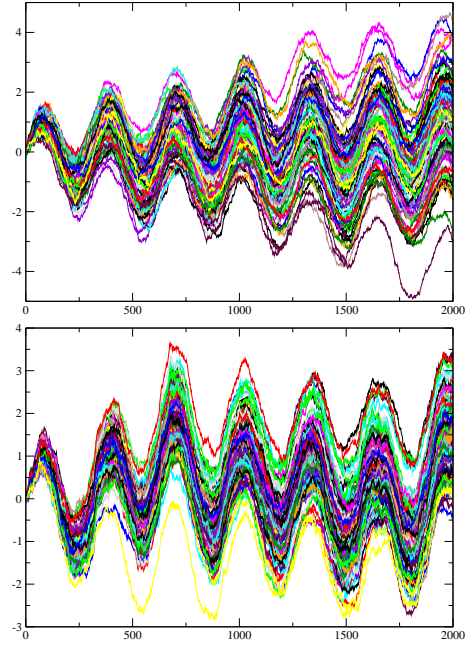

\centering
\begin{tabular}{c}
\epsfig{file=figureSI4.eps,width=0.7\linewidth,clip=} \\
\epsfig{file=figureSI5.eps,width=0.7\linewidth,clip=}
\end{tabular}
\label{fig:exampleRW}
\caption{(A) $N$ Uncorrelated random walkers ($N=100$, $\alpha^2M=0$). (B) Random
 walkers with correlations ($\alpha^2M=10$).}
\end{figure}

In this simple case the exact result is given by $w(t)=F(t)$, $a_i=1$,
and $f_i^{int}(t)=\xi_i(t)$. The important condition for the validity
of the method is given by $A_iA_j\gg\langle G_iG_j\rangle$ and 
is given here by
\begin{equation}
1\gg \alpha^2M
\end{equation}

For $M=0$, the random noises are independent and our method is
very accurate as shown in the main text. 

More generally, in order to assess quantitatively the efficiency of the method, we
compute the Pearson correlation coefficient between the exact
$f^{int}_i(t)$ and the estimate $g_i$ computed with the method. We
plot in figure~$13$ this coefficient versus $\alpha^2 M$.
\begin{figure}[!htp]
\centering
\begin{tabular}{c}
\epsfig{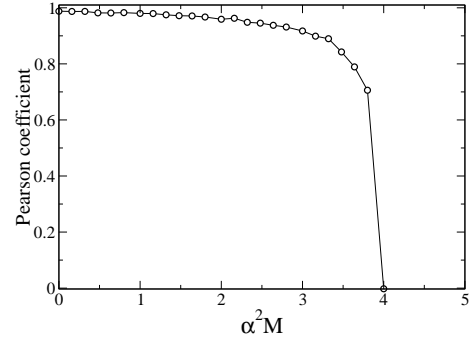}
\end{tabular}
\label{fig:pearson}
\caption{Pearson correlation coefficient between the exact local
  contribution and the local contribution computed with our method
  computed for different values of the correlation ($N=100$, results
  averaged over $100$ realizations).}
\end{figure}
This figure confirms the fact that our method is valid and very
precise provided that the correlations between local contributions are
not too large (here $\alpha^2 M<4$).

\section{ Supporting information 4. Dependence of the $a_i$ on the time interval}

We can compute the quantities $a_i$ for the interval $[t_0,t]$ and by
letting $t$ vary. We then obtain for the crime in the US (in the case
of the crime rates in France, the dataset is not large enough) the
figure~$14$(A).
\begin{figure}[!htp]
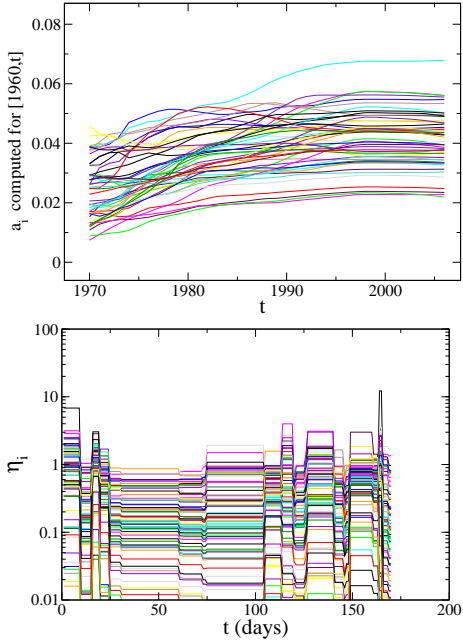

\begin{tabular}{c}
\epsfig{file=figureSI7.eps,width=0.7\linewidth,clip=}\\
\epsfig{file=figureSI8.eps,width=0.7\linewidth,clip=}
\end{tabular}
\label{fig:ait}
\caption{(A) Coefficients $a_i$ computed in the case of US crime for
  the interval $[1960,t]$ with varying $t$ (in years). (B) Coefficients
  $\eta_i$ computed for the SP500 in the interval $[0,125+t]$ ($t$ is
  in days in this case).}
\end{figure}
This figure shows that in the case of the crime rate in the US, the
$a_i$ converge to a stationary value, independent of the time
interval, provided it is large enough. Our method will then lead to
reliable results constant in time.

We also tested our method on the financial time series given by the
500 most important stocks in the US economy [1], and which
composition leads to the $S\&P\;500$ index. Here the `local' units are
the individual stocks ($i=1,...,N=500$), and the (naive) average -
analogue to a national average - is precisely the $S\&P\; 500$ index
time serie. We study the time series for these stocks on the $252$
days of the period $10/2007-10/2008$ and we compute the global pattern
$w(t)$, the coefficients $a_i$, and the parameters $\eta_i$ (defined
in the text) computed for the time window $[10/2007,t]$ for $t$
varying from $04/2008$ to $09/2009$. These quantities $\eta_i$ measure
quantitatively the importance of local versus external fluctuations
for the stock $i$. The results for the $\eta_i$'s are shown in
figure~$6$(B) and display large variations, particularly when we
approach October, 2008, a period of financial crisis. It is therefore not
completely surprising that the $\eta_i$ (and the $a_i$'s) in this case
fluctuate a lot. In some sense, we can conclude that the $a_i$'s
correspond to an average susceptibility to the global trend, are not
invariable quantities and can vary for different periods. We thus see
on this example, that it is important to check the stability of the
coefficients $a_i$ which is an crucial assumption in our method. The
variations of these coefficients is however interesting and further
studies are needed in order to understand these variations.

[1]  Historical Data for $S\& P\;500$ stocks  \url{http://biz.swcp.com/stocks/}

\section{Supporting information 5. Obesity in the US: variances for
  the external and internal contribution}

For the obesity rate series, we compare the variances of the 
internal ($f_i^{int}$) and the external ($a_iw$) contributions. We observe
on the figure $15$ that the variance of the external contribution 
became dominant after the year $\approx 2000$.
\begin{figure}[!h]
\begin{tabular}{c}
\epsfig{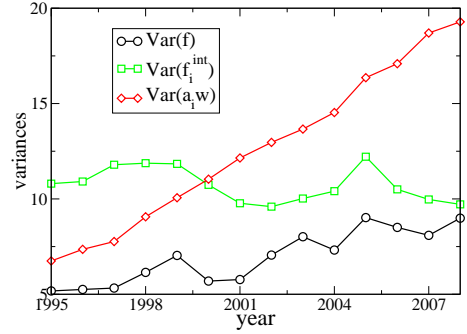}\\
\end{tabular}
\label{fig:varob}
\caption{ Comparison of internal and external fluctuations for the
  obesity in the US. We represent the total variance of the signal
  ($f$), the external ($a_iw$) and the internal contribution ($f_i^{int}$). We
  observe that for the external contribution is dominating since the
  year $2000$. }
\end{figure}

\end{document}